\documentclass[10pt,journal,compsoc]{IEEEtran}
\IEEEoverridecommandlockouts

\ifCLASSOPTIONcompsoc
  \usepackage[nocompress]{cite}
\else
  \usepackage{cite}
\fi

\usepackage{amsmath,amssymb,amsfonts}
\usepackage{algorithmic}
\usepackage{graphicx}
\usepackage{textcomp}
\usepackage[table]{xcolor}
\def\BibTeX{{\rm B\kern-.05em{\sc i\kern-.025em b}\kern-.08em
    T\kern-.1667em\lower.7ex\hbox{E}\kern-.125emX}}

\usepackage{array}
\usepackage{hyperref}
\usepackage{cleveref}
\usepackage[linesnumbered,ruled,vlined]{algorithm2e}
\usepackage{url}

\usepackage{mathtools}
\DeclarePairedDelimiter\ceil{\lceil}{\rceil}

\SetKwInOut{Input}{input}
\SetKwInOut{Output}{output}
\SetNlSty{texttt}{\color{red}}{}
\SetFuncSty{textit}
\SetAlFnt{\footnotesize}
\SetAlgoNlRelativeSize{-2}
\SetKwComment{tcp}{\color{gray}// }{}

\begin{document}

\title{Performance Modeling of Metric-Based Serverless Computing Platforms
}


\author{Nima~Mahmoudi,~\IEEEmembership{Graduate Student Member,~IEEE,}
        and~Hamzeh~Khazaei,~\IEEEmembership{Member,~IEEE}
\IEEEcompsocitemizethanks{\IEEEcompsocthanksitem N. Mahmoudi is with the Department
of Electrical and Computer Engineering, University of Alberta, Edmonton,
Alberta, Canada.\protect\\
E-mail: nmahmoud@ualberta.ca
\IEEEcompsocthanksitem H. Khazaei is with the Department of Electrical Engineering and Computer Science,
York University, Toronto, Ontario, Canada.\protect\\
E-mail: hkh@yorku.ca}
\thanks{Manuscript received April x, 2020; revised August xx, 2020. DOI (identifier): 10.1109/TCC.2022.xxxxxxx}}


\markboth{IEEE Transactions on Cloud Computing,~Vol.~xx, No.~x, February~2022}%
{Mahmoudi \MakeLowercase{\textit{et al.}}: Analysis of Metric-Based Autoscaling in Serverless Computing}

\IEEEtitleabstractindextext{%
\begin{abstract}
Analytical performance models are very effective in ensuring the quality of service and cost of service deployment remain desirable under different conditions and workloads. While various analytical performance models have been proposed for previous paradigms in cloud computing, serverless computing lacks such models that can provide developers with performance guarantees. Besides, most serverless computing platforms still require developers' input to specify the configuration for their deployment that could affect both the performance and cost of their deployment, without providing them with any direct and immediate feedback. In previous studies, we built such performance models for steady-state and transient analysis of scale-per-request serverless computing platforms (e.g., AWS Lambda, Azure Functions, Google Cloud Functions) that could give developers immediate feedback about the quality of service and cost of their deployments. In this work, we aim to develop analytical performance models for latest trend in serverless computing platforms that use concurrency value and the rate of requests per second for autoscaling decisions. Examples of such serverless computing platforms are Knative and Google Cloud Run (a managed Knative service by Google). The proposed performance model can help developers and providers predict the performance and cost of deployments with different configurations which could help them tune the configuration toward the best outcome. We validate the applicability and accuracy of the proposed performance model by extensive real-world experimentation on Knative and show that our performance model is able to accurately predict the steady-state characteristics of a given workload with minimal amount of data collection.
\end{abstract}

}

\maketitle

\IEEEdisplaynontitleabstractindextext

\section{Introduction}
\label{sec:intro}

\IEEEPARstart{S}{erverless} computing platforms are the latest paradigm in the cloud computing
era that aim to minimize the administration tasks required to deploy a workload to the cloud.
They provide developers, software owners, and online services with services like handling
system administration tasks, improving resource utilization, usage-based billing, improved
energy efficiency, and more straightforward application development~\cite{awsserverless, jonas2019cloud}.

Despite having a much faster startup time compared with VM-based deployments, serverless
offerings have shown to lack predictability in key performance metrics. This has rendered them
as unacceptable for many customer-facing products~\cite{jonas2019cloud}.
The issue is exacerbated by the fact that current generation of serverless computing platforms
are workload-agnostic; i.e., using the same management policies for all types of workload with
different needs~\cite{wang2018peeking,shahrad2020serverless,van2018spec}.
This gives us a plethora of possible savings in terms of infrastructure cost and energy consumption
while improving the overall performance by adapting the platform to the unique needs of
each workload~\cite{khazaei2012fine}.

An accurate performance model like the one suggested in this work can benefit both serverless
providers and application developers. Application developers can leverage performance models to
predict the quality of service of their application with different configurations and the respective
cost implications, helping them select the configurations that fits their needs. They also can
use the performance model to find the limitations of their system and plan ahead for large uptakes
in the workload intensity. On the other hand, an accurate performance model can help serverless
providers perform capacity planning and give application developers an estimate on cost and
performance implications of their workload configurations.

A proper performance model for serverless computing platforms should remain tractable while
covering a large portion of the system configuration space. In previous
studies~\cite{mahmoudi2020tccserverless,mahmoudi2020tempperf}, we designed such performance models 
for serverless computing platforms that use scale-per-request autoscaling paradigm predicting
both transient and steady-state quality of service characteristics. In this work, we aim to
develop and evaluate a performance model that captures the unique structure and characteristics
of the most recent paradigm in serverless computing platforms which leverage concurrency value~\cite{mahmoudi2020tccserverless} and 
other metrics to drive autoscaling. The most important examples of these serverless computing platforms
are Knative and Google Cloud Run (which is a managed Knative offering from Google Cloud Platform).

The analytical performance model presented in this work assumes a Poisson arrival process
to address customer-facing open networks which comprise the majority of services which require
strong quality of service guarantees. It has been shown that the arrival process can adequately
be modelled as a Poisson process when there are a large number of clients with each having a
low probability to submit a request at any given
time~\cite{grimmett2001probability,trivedi2010poisson,romero2020infaas,trivedi2018poisson,trivedi2017poisson}.
We impose no restrictions on the service time distribution or service policies by using
data-driven techniques that help extract the unique characteristics of a given workload.
The presented model in this work is highly scalable and can handle a high degree of parallelization
required in large-scale systems. The presented model can help predict the cost and main quality of
service indicators for a given workload, e.g., the average response time.
In addition, the presented performance model can help developers by predicting
the inherent performance-cost tradeoffs for different workload configurations.

The proposed performance model has been validated by extensive experimentation on Knative deployed on our private
cloud computing infrastructure and works with any workload that can be deployed as Docker containers
and accepts HTTP requests.
The development of the model requires a minimal data collection on the target platform to capture the
resource needs of the workload and the effect of concurrency on the quality of service metrics.

The remainder of the paper is organized as follows:
\Cref{sec:system-description} describes the system represented by the analytical performance
model proposed in this work. \Cref{sec:analytical-model} outlines the proposed analytical model.
In \Cref{sec:exp-val}, we present the experimental validation of the proposed model.
In \Cref{sec:related-work}, we survey the latest related work for serverless computing platforms.
\Cref{sec:threats} discusses the threats to the validity of our experiments.
\Cref{sec:conc} summarizes our findings and concludes the paper.

\section{System Description} \label{sec:system-description}

There is very limited documentation available about the scheduling algorithm used in most
serverless computing platforms that use per-request autoscaling~\cite{mahmoudi2020tccserverless}. As 
a result, previous studies have mostly focused on partially reverse engineering these platforms
by running experiments on them~\cite{wang2018peeking,figiela2018performance,bortolini2019investigating,lloyd2018serverless,shahrad2020serverless}.
However, the most recent trend in serverless computing platforms that use 
metric-based autoscaling, Knative~\cite{knative} and Google Cloud Run~\cite{googlecloudrun} for example, are primarily open-sourced and thus 
we can use their source code to develop accurate performance models without speculations\footnote{Note that metric-based autoscaling precedes serverless computing, but new serverless computing generations use different metrics and measurement methods to drive their autoscaling.}.

\begin{figure}[htbp]
\centerline{\includegraphics[width=.85\columnwidth]{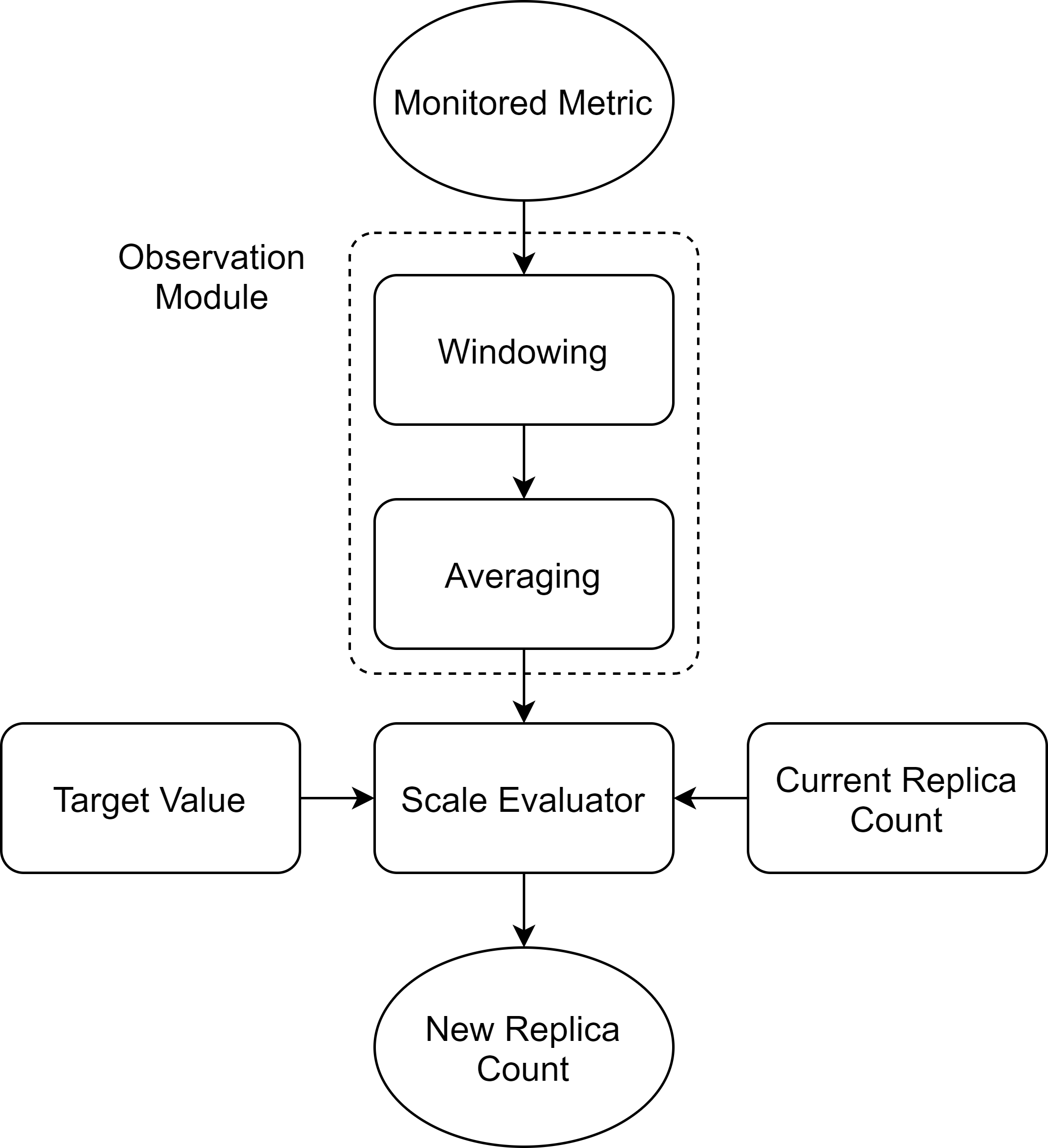}}
\caption{An overview of the Knative scale calculation module. The resulting new replica count will be applied to the cluster.}
\label{fig:knative-autoscale-overview}
\end{figure}

\Cref{fig:knative-autoscale-overview} shows an overview of the Knative scale calculation module.
As shown, we need to choose a monitored metric that will be used to drive the autoscaling in
our deployment. Then, the metric will go through windowing and averaging to generate more stable
observed metrics. 
Using the observed and target values of the used metric, scale evaluator
can calculate the new replica count for a given deployment.
This process is repeated every few seconds to ensure the system remains stable.
In the next sections, we will go through the details of these
steps to outline the system modelled by the proposed performance model.

\subsection{Metrics} \label{sec:system-description-metrics} 

In the metric-based autoscaling approach used in Knative, there are currently two widely
available metrics than can be used to drive autoscaling: 1) Concurrency Value (CC) and 2)
Requests Per Seconds (RPS)~\cite{knativemetrics}. Any of these metrics can be used as the primary monitored metric
and will be compared against the target value for replica count calculations.
These metrics will be monitored by the sidecar container injected by Knative to the Kubernetes
deployment and are collected every second.

\begin{figure}[htbp]
\centerline{\includegraphics[width=1\columnwidth]{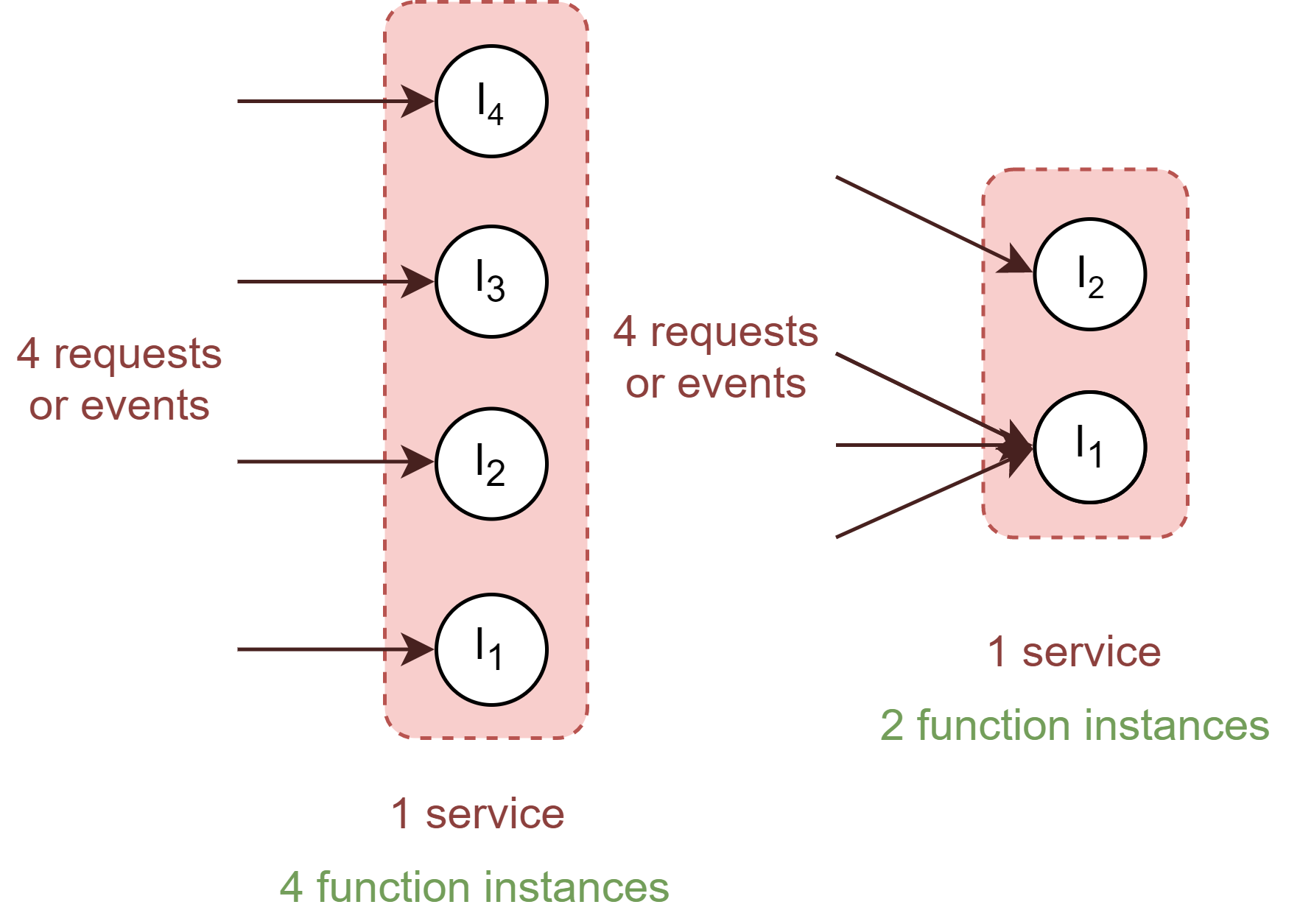}}
\caption{The effect of the concurrency value on the number of function instances needed. The left service allows a maximum of 1 request per instance, while the right service allows a concurrency value of 3.}
\label{fig:concurrency-value}
\end{figure}

\subsubsection{Concurrency Value (CC)}

\begin{figure}[htbp]
\centerline{\includegraphics[width=1\columnwidth]{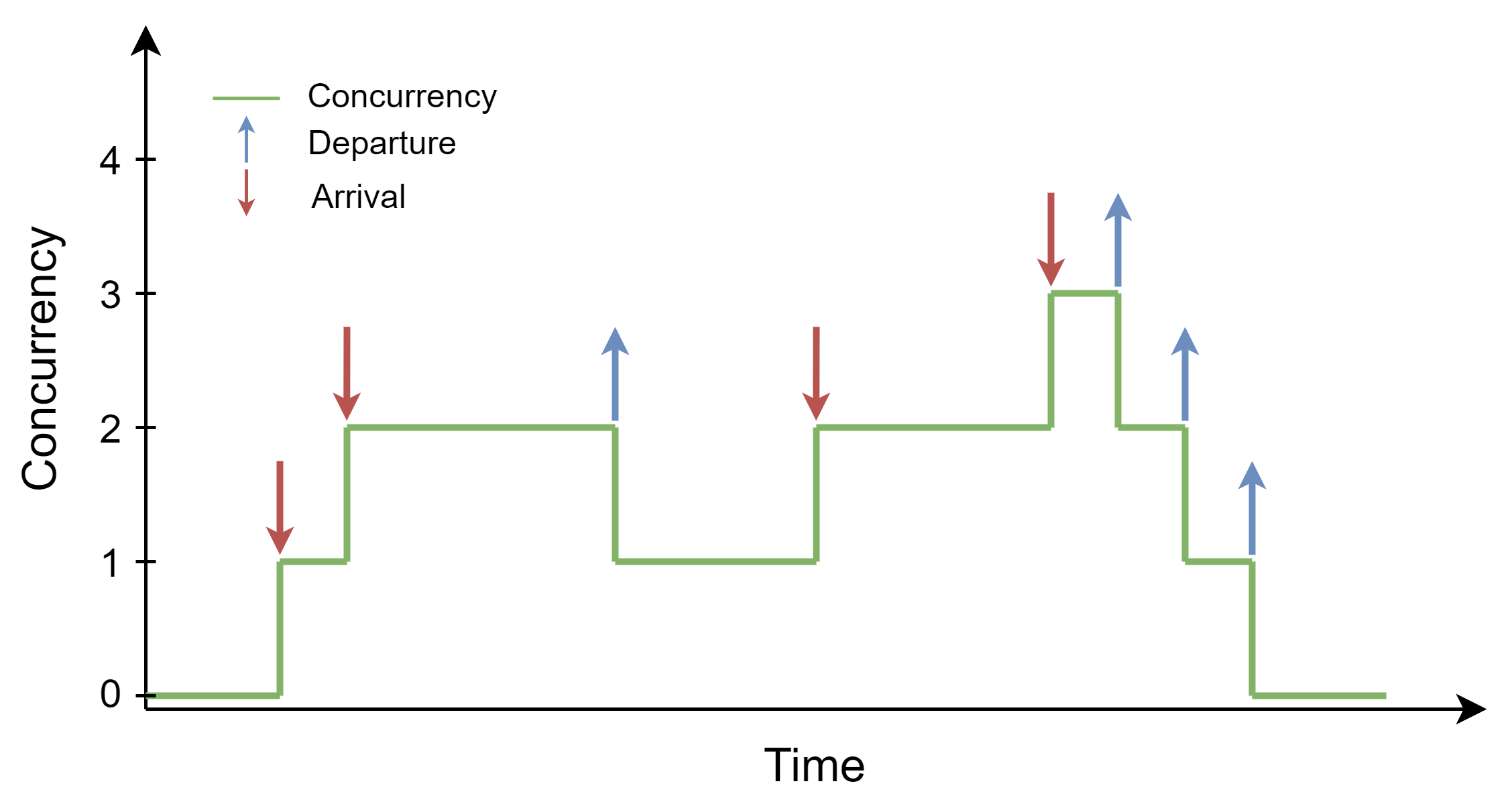}}
\caption{An example scenario of the change in the container concurrency value. The effect of request arrival and departure on concurrency value is shown over time.}
\label{fig:concurrency-example}
\end{figure}

Unlike most public serverless computing platforms that primary use scale-per-request autoscaling
such as AWS Lambda, Google Cloud Functions, Azure Functions, and IBM Cloud Functions, Knative and
consequently Google Cloud Run allow several requests to enter the same function instance at the
same time. The number of concurrent requests being processed by the same container is called
the \textit{concurrency value} in Knative documentations. \Cref{fig:concurrency-value} shows the possible
effect of concurrency in serverless computing platforms which could lead to fewer function instances.
Concurrency value is the default metric used in Knative and is the only metric supported in
Google Cloud Run. Thus, we will focus more on this metric throughout this work, but the
proposed performance model also works with RPS as the monitored metric. 
\Cref{fig:concurrency-example} shows an example of how concurrency changes with request arrival
and departure in each container. As can be seen, any request arrival results in an increment
in the concurrency value and any request departure results in a decrement in the monitored
concurrency value.

\subsubsection{Requests Per Seconds (RPS)}

The arrival rate for each container or RPS is another monitored metric supported by Knative.
However, at the moment this metric is not being supported by Google Cloud Run.
The measurement of this metric is straightforward, the monitoring module monitors
the number of requests arriving to each container every second and reports the resulting value.

\subsection{Observation Module} 
\label{sec:system-description-observation-module}

The observation module is responsible for collecting monitored metrics from all containers,
generating the average values for every second, and calculating the moving average throughout time
according to the the \textit{stable window} configuration. The default value of the \textit{stable window}
is 60 seconds in Knative.

The output of this module is the \textit{observed value} that will be used for driving
scaling decisions. The role of this module is to generate stable observations in order
to avoid making premature decisions in the scaling evaluations.


\subsection{Scale Evaluator Module}

As discussed in \Cref{sec:system-description-observation-module}, the observation module generates
stable averaged measurements from single container measurements of the monitoring module.
The \textit{Scale Evaluator} uses these measurements and the user-specified configurations to 
generate the new replica count ordered by the evaluator in each evaluation using the following equation\footnote{source: \url{https://github.com/knative/serving/blob/master/pkg/autoscaler/scaling/autoscaler.go}. Last accessed 2021-02-01.}:

\begin{equation}
{NewOrderedReplica} = \ceil*{\frac{{ObservedValue}}{{TargetValue}}}
\end{equation}
where the \textit{Observed Value} and \textit{Target Value} are values of the chosen monitoring
metric by the user, i.e., concurrency or RPS. By default, the Knative autoscaling evaluation
takes place every $T_{eva}$ (2 seconds in Knative), setting the new replica target on the Kubernetes deployment.


\section{Analytical Model} \label{sec:analytical-model}

In \Cref{sec:system-description}, we outlined the details of the system modelled by our proposed
performance model. In this section, we will go through the details of the performance models based
on the described system. Our primary focus here is to predict steady-state metrics of a given
workload based on the input system configurations.

\begin{figure}[htbp]
\centerline{\includegraphics[width=0.6\columnwidth]{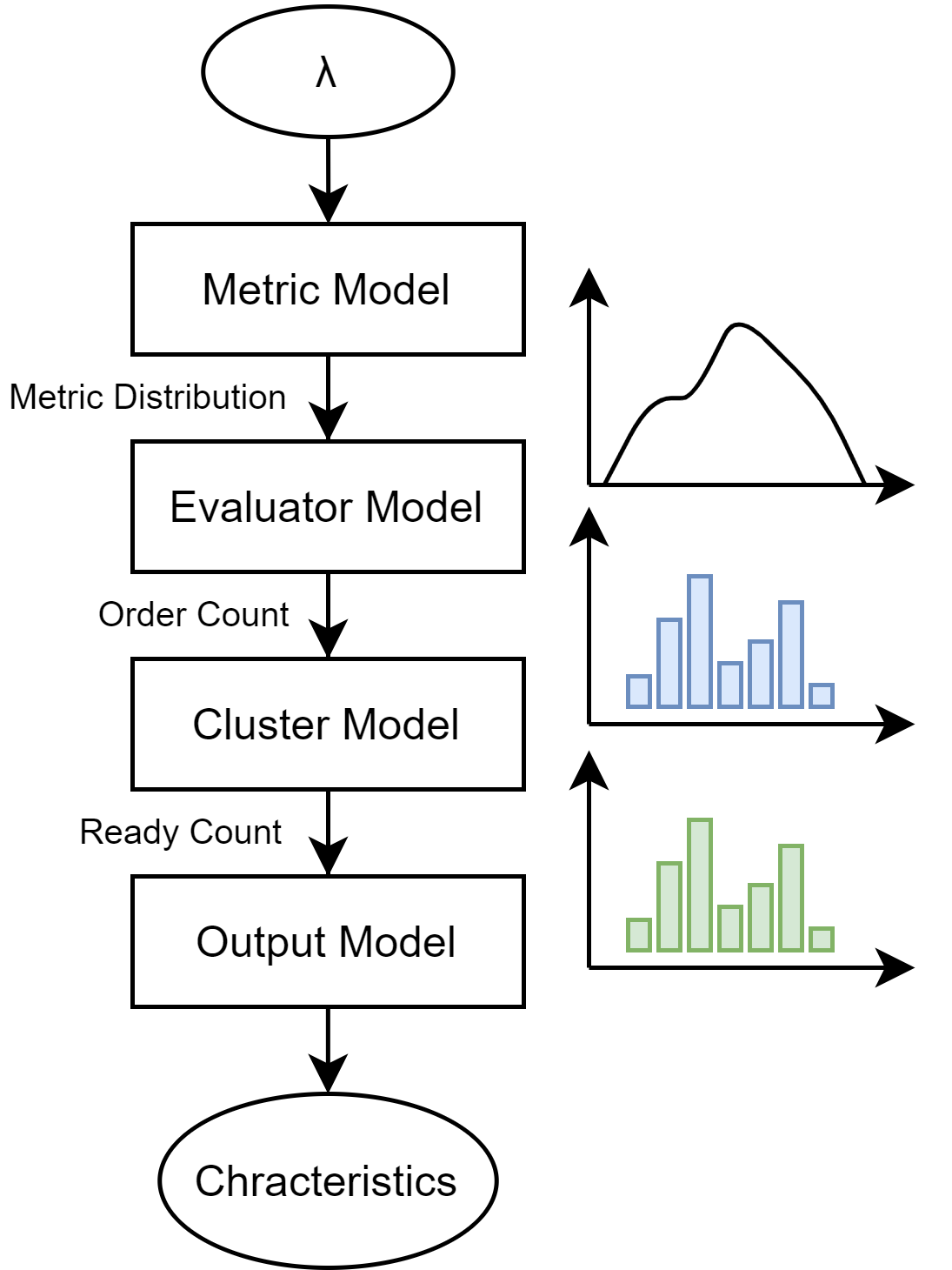}}
\caption{An overview of the proposed performance model.}
\label{fig:performance-model-overview}
\end{figure}

\Cref{fig:performance-model-overview} shows an overview of the proposed performance model.
As can be seen, given the arrival rate, the metric module can use the workload profile
to calculate the distribution of the monitored autoscaling metric (i.e., concurrency value or RPS).
This step is very important as it captures several important characteristics of a given workload
like the amount of work needed for each request and its distribution, along with the deployment
configuration like the CPU and memory configuration of the deployment. This is mainly due
to the fact that the effect of all of the aforementioned properties is captured in the data
achieved from the monitoring module.
Given this value, the evaluator model can estimate the probability of setting different values for
the replica count of the service deployment. 
Having calculated the probability of setting the replica count to different values
and using the estimated provisioning/deprovisioning rates, we can estimate
the probability of seeing different replica counts using the cluster model.
Finally, using the ready container replica
count and by using the output model, we calculate the steady-state estimates for different
characteristics of the deployment.

In the following subsections, we present the calculation of different
parameters in the analytical models using the symbols defined in \Cref{tab:symbols}; we
will elaborate on the details of the aforementioned sub-models.

\begin{table}[!t]
  \renewcommand{\arraystretch}{1.4}
  \caption{Symbols and their corresponding descriptions.}
	\centering
       	\rowcolors{2}{white}{gray!25}
	\begin{tabular} {c|p{6.8cm}}
		\hline
		\textbf{Symbol} & \textbf{Description}	\\
		\hline
		$\lambda$ & Mean arrival rate of requests \\
		$N$ & Number of function instances \\
		$\overline{N}$ & Average replica count \\
		$N_{max}$ & Maximum number of function instances \\
		$N_{ord}$ & Ordered number of function instances \\
		$OV$ & Observed value of the monitored metric \\
		$f_{OV}(\cdot)$ & Density function for the observed value \\
		$F_{OV}(\cdot)$ & Distribution function for the observed value \\
		$\textit{TV}$ & Target value for the monitored metric \\
		$\textit{MM}$ & Metric model \\
		$\textit{EM}$ & Evaluator model \\
		$T_{eva}$ & Time between consecutive evaluations \\
		$Q$ & The CTMC transition rate matrix \\
		$P$ & The DTMC transition probability matrix \\
		$\pi$ & The steady-state distribution \\
		$\mu_{pro}$ & Mean provisioning service rate \\
		$\mu_{dep}$ & Mean deprovisioning service rate \\
		$\overline{RT}$ & Mean Service Response Time \\
		$\overline{RT}_N$ & Average response time with $N$ containers \\
		$\textit{RTF}$ & Response Time Function \\
		$N_{opr}$ & Number of overprovisioned instances \\
		$N_{upr}$ & Number of underprovisioned instances \\
		$\overline{C}$ & Average concurrency level \\
		$C_i$ & Average concurrency for state number $i$ \\
		\hline
		\end{tabular}
	\label{tab:symbols}
\end{table}

\subsection{Metric Model}

As discussed in \Cref{sec:system-description-metrics}, there are two main metrics that can
be used with this family of serverless computing platforms, namely concurrency (CC) and the arrival rate for each container (RPS).
The chosen metric will then be processed by the observation module and will be windowed and
averaged to be used in scaling operation. The goal of the metric model is to estimate
the distribution of the observed values for a given arrival rate. However, different
applications show very different behaviours when processing more than one request.

Processing times, and consequently measured concurrency values, are largely influenced
by factors like service policy (whether the application uses First Come First Serve,
Processor Sharing, or a combination of both) and its reliance on external service.
Intuitively, using a fair load balancer, we can safely assume that
the service time and concurrency value for a given workload largely depend only on arrival rate
per container, i.e., RPS or $\lambda/N$. 
Also, since the observed metric is being averaged over several containers and
over 60 measurements throughout the time, it can safely be assumed to be coming from a
Gaussian distribution due to the central limit theorem.
Thus, we decided to use data-driven methods to estimate
the observed metric average and standard deviation. As a result, we need a few minutes of
data collection for a given workload to build our data-driven model before generating
predictions.
We used 5 minutes of data collection for our experiments and collecting enough data to have at least 100 measurements is suggested to achieve an acceptable accuracy, but gathering more data can always improve the accuracy of the system.

In this step, our goal is to find the function $\textit{MM}$ that estimates the following:

\begin{equation}
f_{OV}(x) \approx \textit{MM}(x; \lambda/N)
\end{equation}
where $f_{OV}(\cdot)$ denotes the observed value density function, $\textit{MM}$ denotes the metric
model, $\lambda$ denotes the arrival rate, and $N$ represents the number of ready containers
in the cluster. Using this distribution, the evaluator model can estimate the number of ordered
containers and their probabilities.
Note that to develop this model, we are assuming a homogeneous cluster where each
container has a similar amount of CPU. We also assume a good performance isolation
between containers which is safe assumption due to the high level of performance
isolation in the modern managed Knative services like Google Cloud Run.

\subsection{Evaluator Model}

The evaluator model has been designed to model the behaviour of the \textit{Scale Evaluator}
module of the autoscaler. In this model, we use the observed value density function
$f_{OV}(\cdot)$ to calculate the probability of different values for the new number of
ordered replica count. This module will take the \textit{Target Value} ($\textit{TV}$),
maximum replica count ($N_{max}$), and other configuration that affect the ordered replica count
(e.g. maximum scale up/down rate) into account. We know from the system description that
the new ordered replica count in each evaluation is given by the following equation:

\begin{equation}
N_{ord} = \ceil*{\frac{\textit{OV}}{\textit{TV}}}
\end{equation}
where $N_{ord}$ is the new number of ordered replica count, $\textit{OV}$ represents
the observed value, and $\textit{TV}$ is the target value set by the user. Thus,
we can calculate the probability of a specific value ($i$) for $N_{ord}$:

\begin{equation}
\begin{split}
    Pr\{N_{ord}=i\} &= Pr\{\ceil*{\frac{\textit{OV}}{\textit{TV}}} = i\} \\
    &= Pr\{(i-1) < \frac{\textit{OV}}{\textit{TV}} \leq i\} \\
    &= Pr\{(i-1) \cdot \textit{TV} < \textit{OV} \leq i \cdot \textit{TV}\} \\
    &= F_{OV}(i \cdot \textit{TV}) - F_{OV}((i-1) \cdot \textit{TV})
\end{split}
\end{equation}
where $F_{OV}(\cdot)$ is the cumulative density function of the observed value
which can be calculated from the metric model using the following:

\begin{equation}
    F_{OV}(x) = Pr\{OV \leq x\} = \int_{-\infty}^{x} f_{OV}(x) \,dx
\end{equation}

Repeating this procedure for any possible number of containers in the range $[1,N_{max}]$,
we get the probability of having different values for the number of ordered instance
counts in a given deployment. 

\begin{equation}
    \textit{EM}(i; f_{OV}) = F_{OV}(i \cdot \textit{TV}) - F_{OV}((i-1) \cdot \textit{TV})
\end{equation}
where $\textit{EM}(i; f_{OV})$ is the probability of setting the ordered replica count to $i$
given $f_{OV}$.
These results help us build a complete and accurate
cluster model to predict the overall behaviour of our deployment.

\begin{figure}[htbp]
\centerline{\includegraphics[width=1\columnwidth]{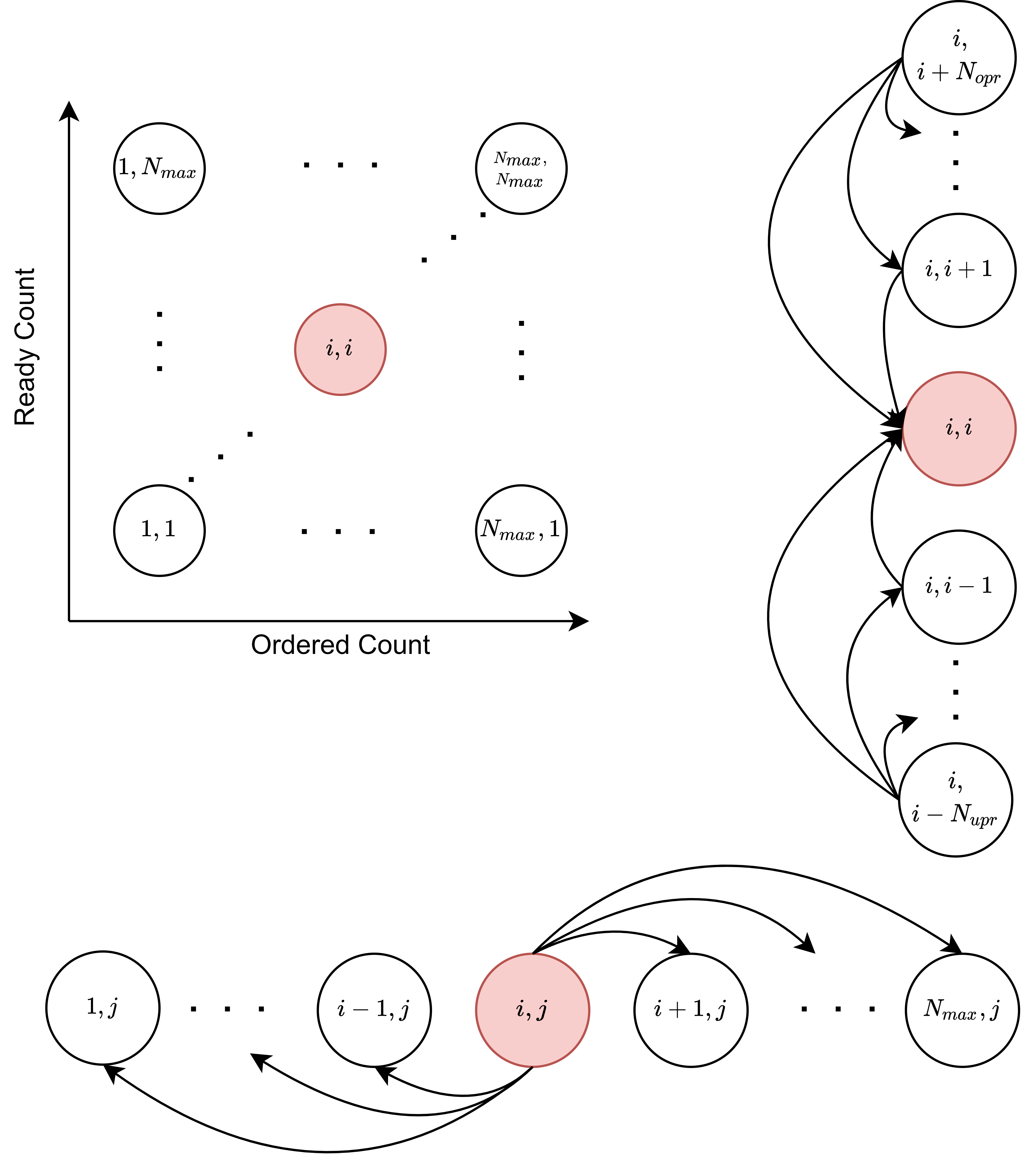}}
\caption{An overview of the proposed cluster model along with its vertical and horizontal components.}
\label{fig:cluster-model-overview}
\end{figure}

\subsection{Cluster Model}

In this section, we will detail the design of the proposed Discrete-Time Markov Chain (DTMC)
representing the status of our metric-based serverless deployment in the cluster.

\begin{figure}[htbp]
\centerline{\includegraphics[width=.7\columnwidth]{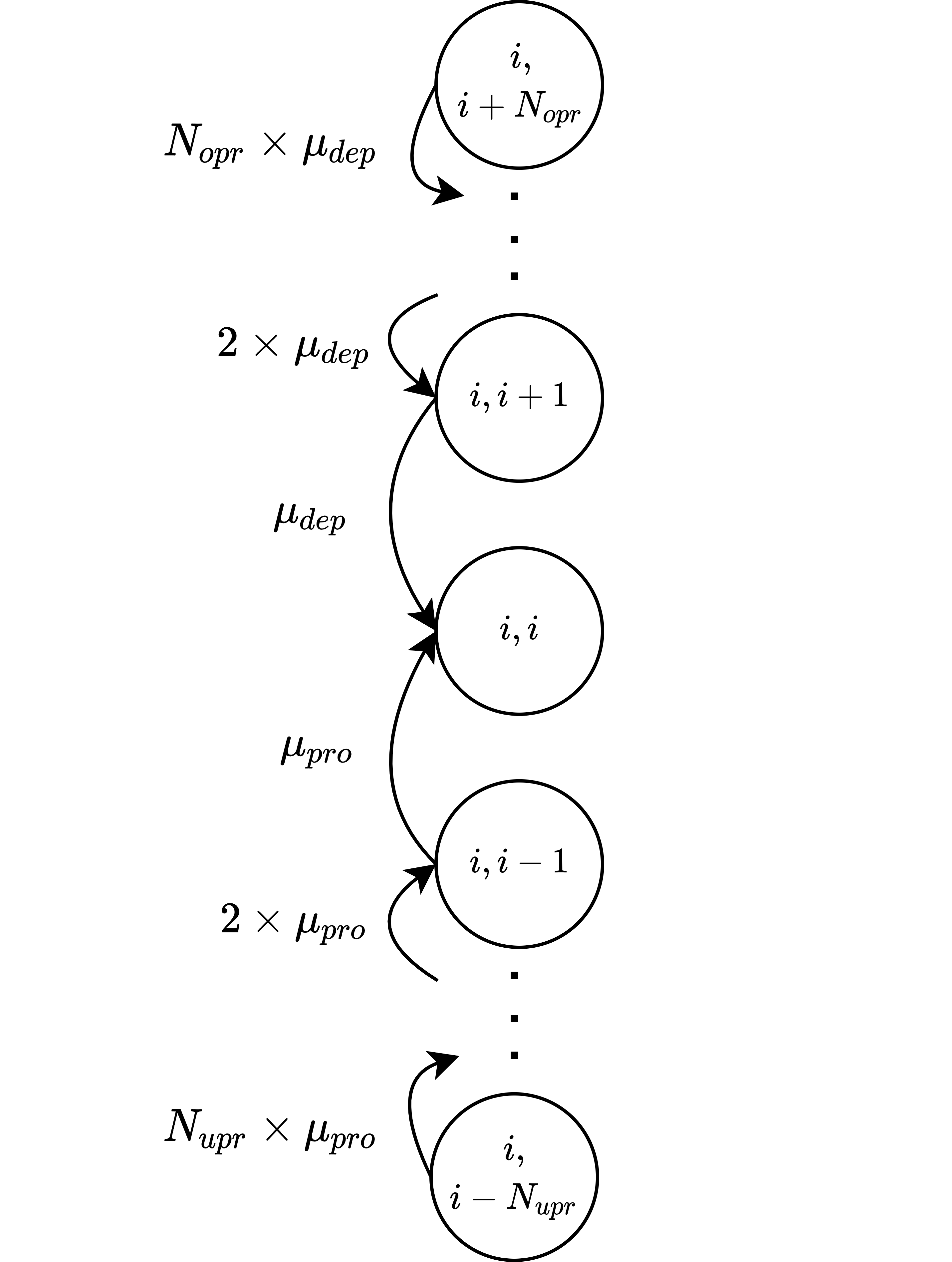}}
\caption{An overview of the underlying infrastructure CTMC model used in the cluster model. $N_{opr}$ and $N_{upr}$ signify the number of overprovisioned and underprovisioned containers and $\mu_{pro}$ and $\mu_{dep}$ represent the provisioning and deprovisioning service rates, respectively.}
\label{fig:cluster-model-infra-ctmc}
\end{figure}

\Cref{fig:cluster-model-overview} shows an overview of the proposed two dimensional
DTMC where the x-axis represents the number of containers ordered by the evaluator
and the y-axis shows the number of containers that are currently in the \textit{Ready}
state and can accept incoming requests. To build the resulting model, we have chosen
to evaluate the system at the moment after each evaluation by the scale evaluator. As a result,
the newly set order count has not had the chance to affect the system yet and thus gives us
the ability to decouple the \textit{single-step} infrastructure effect of provisioning or deprovisioning of containers
from the effect of the execution of the evaluator.
This is due to the fact that we are modelling a physical system here, and like any other
physical systems, configuration changes cannot affect the system instantly and require
some time to do so. It is worth noting that our model still captures the relationship between
the number of ordered containers and ready containers via vertical transitions
(better shown in \Cref{fig:cluster-model-infra-ctmc}) in consequent steps
of the model.

There are two main forces causing the change in the system:
1) change in the order count due to execution of the scale evaluator; and 2) change
in the number of deployment containers due to provisioning or deprovisioning of containers.
Due to the aforementioned decoupling between these two forces that affect our state,
they will be independent and thus any transition probability in these two dimensions
can be broken down as the following:

\begin{equation}
    P_{(i,j),(i',j')} = P_{i,i'}(j) \times P_{j,j'}(i)
\end{equation}
where $P_{(i,j),(i',j')}$ is the probability of transitioning to state $(i',j')$ given
our current state is $(i,j)$, $P_{i,i'}(j)$ is the probability of transitioning from
column $i$ to column $i'$ from row $j$, and $P_{j,j'}(i)$ is the probability of
transitioning from row $j$ to row $j'$ from column $i$. As can be seen, $P_{i,i'}(j)$
does not depend on $j'$ and $P_{j,j'}(i)$ does not depend on $i'$, which significantly reduces
the computational complexity of the overall performance model.

To calculate $P_{i,i'}(j)$, we use the evaluator model developed in the previous section.
We assume the result of each scale evaluation is independent from previous evaluations
and thus we have:

\begin{equation} \label{eq:state-trans-prob-new-order}
\begin{split}
    f_{OV}(x) &= \textit{MM}(x; \lambda/j) \\
    P_{i,i'}(j) &= \textit{EM}(i'; f_{OV})
\end{split}
\end{equation}

To obtain $P_{j,j'}(i)$, we need to analyze how the infrastructure reacts when provisioning
or deprovisioning of containers for a given deployment takes effect. To do so, we use the Continuous-Time
Markov Chain (CTMC) model shown in \Cref{fig:cluster-model-infra-ctmc} and solve for
possible transitions after $T_{eva}$ units of time. In this model, we assume
exponentially distributed service times for provisioning/deprovisioning 
for which the rate is proportional to the amount of the underlying resources. As a result, $P_{j,j'}(i)$ becomes the probability of starting
in state $(i,j)$ and provisioning/deprovisioning enough containers to get to $j'$
containers in the cluster after $T_{eva}$ units of time. 

To solve the resulting CTMC model,
we use the one-step transition rate matrix $Q$ to get the state distribution
$\pi'$. In this matrix, each element
located in row $x$ and column $y$ shows the transition rate at which we transition
from state $x$ to state $y$. Diagonal elements are defined in a way to satisfy 
$Q_{x,x} = -\sum_{y\neq x} Q_{x,y}$. To solve the resulting CTMC, we have to solve
the following equation:

\begin{equation}
    \frac{d\pi'}{dt} = \pi' Q \Rightarrow \pi'(t) = \pi'(0) e^{Qt}
\end{equation}
\noindent
which can be calculated using the method proposed by Al-Mohy et al.~\cite{al2010new}.

Using the state distribution $\pi'$, we can calculate the transition probabilities $P_{j,j'}(i)$ using the following equation:

\begin{equation} \label{eq:state-trans-prob-infra}
    P_{j,j'}(i) = \pi'_{j'}(T_{eva})
\end{equation}

Using \Cref{eq:state-trans-prob-new-order,eq:state-trans-prob-infra}, we can build the
transition probability matrix $P$ for the cluster model shown in \Cref{fig:cluster-model-overview}.
To analyze the steady-state behaviour of the system, we need to calculated the
limiting probability $\pi_s$ for any state $s$ where~\cite{harchol2013performance}:

\begin{equation}
    \pi_s = \lim_{n\to\infty}{P_{s,s'}^n}
\end{equation}
where $\pi_s$ is the probability that chain is in state $s$, independent of the starting
state $s'$. Using these limiting probabilities, we can calculate the limiting distribution
$\pi$:

\begin{equation}
    \pi = (\pi_1,...,\pi_{M}), \sum_{x=1}^{M}{\pi_x} = 1
\end{equation}
where $M$ signifies the total number of states, which is $M=N_{max}^2$ here.
It can be shown that the resulting limiting distribution is $\pi$ if
$\pi P=\pi$ and $\sum_{x=1}^{M}{\pi_x} = 1$. This system of equations can be solved
using the method outlined in~\cite{hermansolvedtmc}.



After knowing the steady-state probability of being in each state via $\pi$,
we need to calculate different desired metrics and characteristics of the workload.

\subsection{Output Model}
In the previous section, we went over the details of cluster model, which is
used to calculate the limiting distribution. In this section, we will use the resulting
state distribution to calculate metrics of interest in a given Knative deployment.
Two of the most important metrics in a given deployment are \textit{average response time}
as an indicator for Quality of Service (QoS), \textit{average replica count} as an
indicator for cost, and \textit{average concurrency} as a metric used in infrastructure
planning like database capacity planning, etc.
Here, we will go over the details of calculating each of these metrics.

\subsubsection{Average Response Time}

Average response time is one of the
most widely used metrics to indicate the quality of
service for a given deployment in the context of web services.

Intuitively, assuming negligible overhead in the Kubernetes routing mechanism
(compared to the request processing time), the average response time for a given
workload is only a function of arrival rate per container ($\lambda/N$), or in other
words the amount of work given to each containers. However, this relationship
is highly dependent on the type of workload, its parallel or concurrency features,
and the type of workload being used (CPU, I/O, or memory intensive, or a combination
of them).

As a result, we have decided to use automated data-driven methods to extract to
which extent does the average response time rely on the arrival rate per container and
show the result as the following:

\begin{equation}
    \overline{RT}_N = \textit{RTF}(\lambda / N)
\end{equation}
where $\overline{RT}_N$ is the average response time of the service when we
have $N$ containers and RTF shows the response time function, estimated using
regression methods from our brief profiling window.
To calculate the total average response time, we use the state probabilities
calculated:

\begin{equation}
\begin{split}
    \overline{RT} &= \sum_{i=1}^{M} \pi_i \overline{RT}_{N_i} \\
    &= \sum_{i=1}^{M} \pi_i \textit{RTF}(\lambda / N_i)
\end{split}
\end{equation}
where $M$ is the number of states, $N_i$ is the number of ready containers in
state number $i$, and $\pi_i$ is the probability of being in state number $i$
at any time step.

\subsubsection{Average Replica Count}

Nowadays elite cloud vendors use very complicated and regularly changing pricing schema with multitude of
charges for different services and providing a complete pricing model for them is infeasible.
However, there are mainly two sets of factors used in calculating the incurred cost of a given deployment
in a serverless setting: 1) per-request costs and 2) per-instance cost. For a given arrival
rate, the calculation of per-request costs are rather straightforward since we have an estimate
of $\lambda \cdot T$
for the number of requests in any given time window with length $T$. 
However, calculating per-instance
costs relies on the system configurations and characteristics and can vary drastically
based on these settings. To provide application developers and operations experts with
a tool that helps them understand the tradeoffs of their deployments, we leverage the
developed performance model to calculate the average number of running instances in
the cluster.

To calculate the average replica count, we can use the state probabilities calculated
in previous sections:

\begin{equation}
    \overline{N} = \sum_{i=1}^{M} \pi_i N_i
\end{equation}
where $\overline{N}$ is the average replica count and $N_i$ is the number of
ready containers in state number $i$.

\subsubsection{Average Concurrency}

The average concurrency level per container is a measure that can help application
developers set reasonable resource limits and configurations for a given service
as well as tune other services they rely upon, e.g., databases. The average concurrency
level ($\overline{C}$) can also be calculated using state probabilities:

\begin{equation}
    \overline{C} = \sum_{i=1}^{M} \pi_i C_i
\end{equation}
where $\overline{C}$ is the overall average concurrency and $C_i$ is the average concurrency
for state number $i$. To get $C_i$, we can use the metric model for concurrency value:

\begin{equation}
    C_i = \int_{0}^{\infty} x \cdot \textit{MM}(x; \lambda / N_i) dx
\end{equation}

\section{Experimental Evaluation} \label{sec:exp-val}

In this section, we introduce our evaluation of the proposed analytical
performance model using experimentation on our Knative installation.
The code for performing and analyzing the experiments used in this section
can be found in our public GitHub repository\footnote{\url{https://github.com/pacslab/conc-value-perf-modelling}},
along with installation and deployment instructions of various workloads used in
this study.
To the best of authors' knowledge, no other work has proposed a performance model for this type of serverless computing platforms. As a result, our experimental results only include our measurements compared to the proposed performance model.

\begin{table}[htbp]
\renewcommand{\arraystretch}{1.4}
\begin{center}
    \centering
    \rowcolors{2}{white}{gray!25}
    \caption{Configuration of the VMs in the experiments.}
    \label{tab:vm-config}
    \begin{tabular}{l  p{15em}} 
        \hline
        \bf Property & \bf Value\\
        \hline
        vCPU & 4 \\ 
        \hline
        RAM & 8GB \\ 
        \hline
        HDD & 40GB \\ 
        \hline
        Network & 1000Mb/s \\ 
        \hline
        OS & Ubuntu 20.04 \\ 
        \hline
        Latency & \textless 1ms \\ 
        \hline
    \end{tabular}
\end{center}
\end{table}

\subsection{Experimental Setup}

To perform our experiments, we used 4 Virtual Machines (VMs) on the 
Cybera Cloud~\cite{cybera} with the configuration shown in \Cref{tab:vm-config}.
Of the VMs used, 3 joined in a Kubernetes cluster and 1 used as the client.
We found the cluster size sufficient for our experiments due to the fact that
modern application architectures include several smaller deployments each
receiving a portion of the traffic and our approach aims to model these individual deployments.
For our cluster, we used Kubernetes version \textit{1.20.0} with 
Kubernetes client (kubectl) version \textit{1.18.0}. For the client, we used
\textit{Python 3.8.5}. To generate client requests based on a Poisson process,
we used our in-house workload generation library
\footnote{\url{https://github.com/pacslab/pacswg}}
which is publicly available through PyPi\footnote{\url{https://pypi.org/project/pacswg}}.
The result is stored in a CSV file and then processed using
Pandas, Numpy, Matplotlib, and Seaborn.
The dataset, parser, and the code for extraction of system parameters and properties are also
publicly available in the project's GitHub repository.
For all experiments, we performed the experiment in 6
batches totalling one hour for each combination of configurations to get accurate results.
Based on the tests on our cluster, we used the estimated values of $\mu_{pro}=1$ and
$\mu_{dep}=2$ events per second.


\subsection{Workloads}

To evaluate the proposed performance model, we used workloads in Python and Go programming languages to represent different types of applications. The results for all of these workloads can be found on the project's GitHub repository. 
To improve the generalizability of the results, these workloads each include several parts designed to dominate one or more resources, and by using different combinations of these workloads, we can represent a large spectrum of different workloads.
We also included scripts that automate the process of deployment, load testing, and logging of
the results. We present a representative subset of these results here due to space limitation.
For \textit{workload 1}, we used the work of Wang et al.~\cite{wang2018peeking} written in 
\textit{Python} with minor modifications and utilizing Flask as the web server.
This workload is a combination of CPU intensive and I/O intensive workloads.
For \textit{workload 2}, we used a standard and open-source suite of benchmarks implemented by the Knative community in the Go programming language\footnote{For more information, visit \url{https://knative.dev/docs/serving/autoscaling/autoscale-go/}}.

The regression method is not an integral part of our performance model and thus any
regression method with enough accuracy for a given workload can be used.
To predict the mean concurrency value based on $\lambda / N$ for experimental workloads, we used a simple polynomial
regression of the following form with no training on the intercept:

\begin{equation}
    y = \alpha_1 \cdot x + \alpha_2 \cdot x^2
\end{equation}
where $\alpha_i$s are the trained parameters of the model, $y$ represents the output value, and $x$ represent the input to the model ($\lambda / N$).
This method has a low number of parameters, which increases its interpretability. Besides, its variance is low which enables us to train it accurately using a limited amount of data. It also allows us to control the regression's behaviour in extreme values to make sure it presents sensible values for the model. For example, in very low arrival rates, we know the measured concurrency should approach zero, which is integrated into this model.
In our experiments with \textit{workload 1} and \textit{workload 2}, the resulting fit had a Mean Squared Error (MSE) of $0.1004$ and $0.004$ and $R^2$ score of $0.9875$ and $0.9991$, respectively.


Similarly and for the same reasons, we used a polynomial regression but this time with an intercept to get the response
time from average arrival rate per container. The resulting function is of the following
form:

\begin{equation}
    y = \alpha_0 + \alpha_1 \cdot x + \alpha_2 \cdot x^2
\end{equation}
where $\alpha_i$s are the trained parameters of the model, $y$ represents the output, and $x$ represent the input to the model ($\lambda / N$). One nice feature that can be enforced with a simpler
regression method like the one presented here is that we can control it to approach the
service time of the workload when arrival rate per container approaches zero.
In our experiments with \textit{workload 1} and \textit{workload 2}, the resulting fit had a Mean Squared Error (MSE) of $0.0380$ and $8.5177*10^{-7}$ and $R^2$ score of $0.8159$ and $0.6259$, respectively.


\subsection{Experimental Results} \label{sec:exp-results}

In this section, we go through our experimental results and their predicted counterparts.
To get the results for each point shown in the experimental plots, we ran a test with a specific Poisson arrival process for every single point; we also eliminate the first 5 minutes of the experiment to eliminate the transient effect (i.e., warm-up effect).

\begin{figure}[htbp]
\centerline{\includegraphics[width=1\columnwidth]{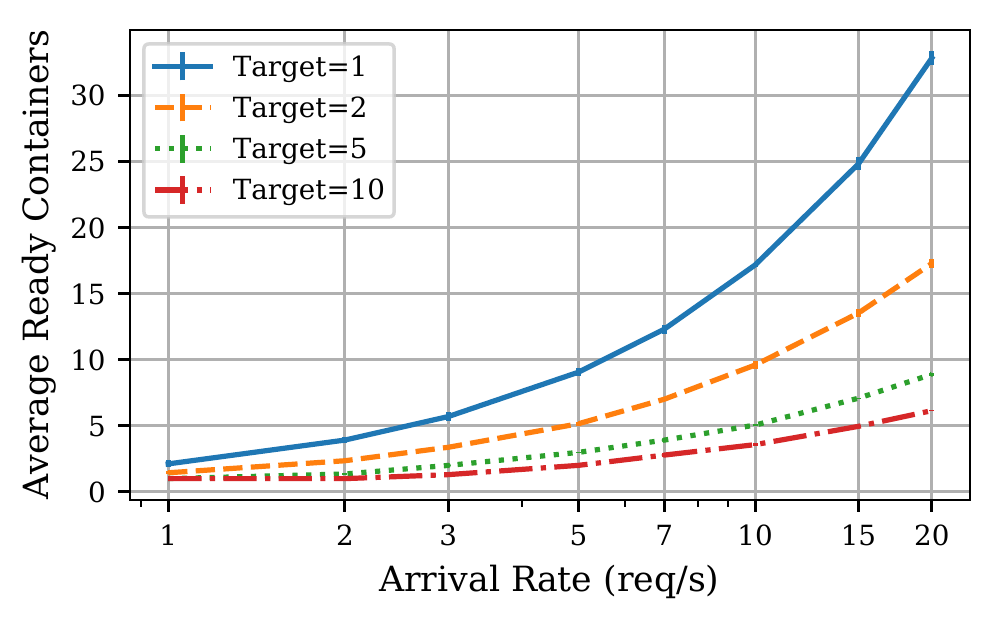}}
\caption{The measured average number of containers ready to server requests versus the fixed arrival rate for different target concurrency values in our experiments. Note that the x-axis is on a logarithmic scale. The vertical bar shows the 95\% confidence intervals which in this case were very small because experiments were long enough to have very accurate results.}
\label{fig:expres01}
\end{figure}

\begin{figure}[htbp]
\centerline{\includegraphics[width=1\columnwidth]{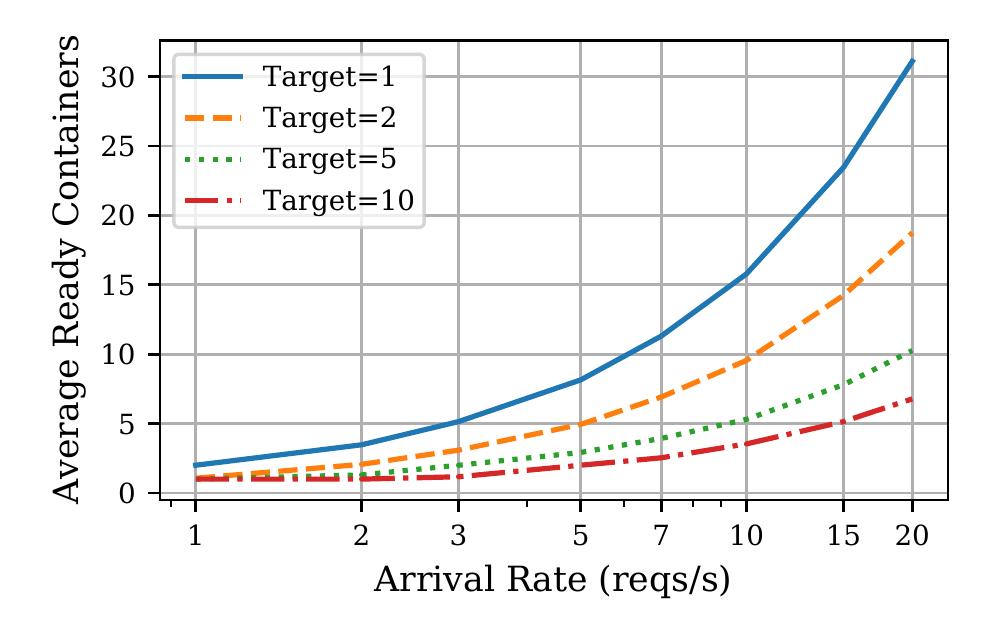}}
\caption{The predicted average number of containers ready to server requests versus the fixed arrival rate for different target concurrency values. Note that the x-axis is on a logarithmic scale.}
\label{fig:expres01_model}
\end{figure}

\begin{figure}[htbp]
\centerline{\includegraphics[width=1\columnwidth]{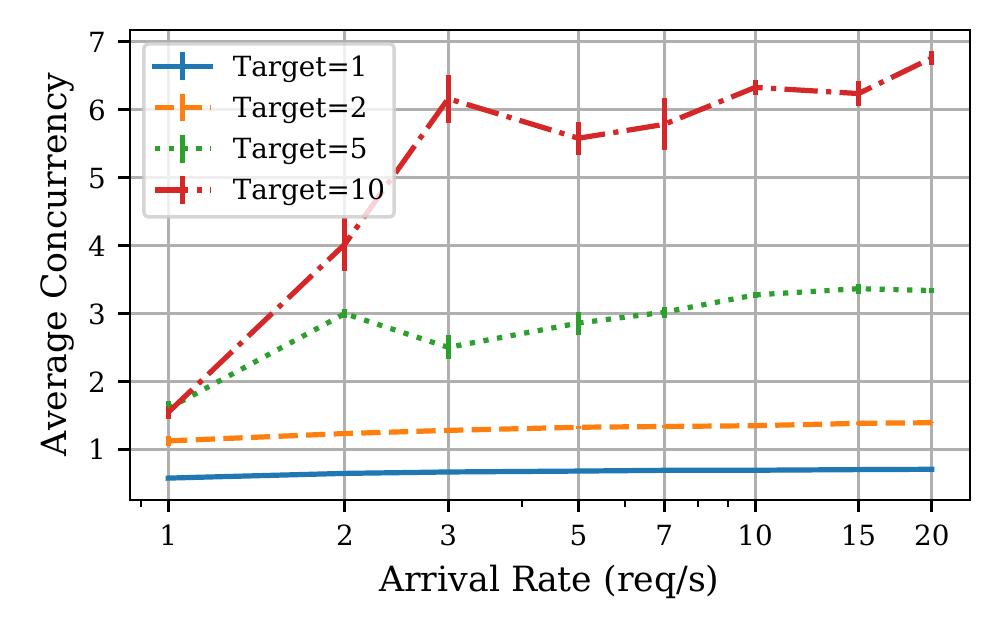}}
\caption{The measured average concurrency value versus the fixed arrival rate for different target concurrency values in our experiments. Note that the x-axis is on a logarithmic scale. The vertical bar shows the 95\% confidence intervals.}
\label{fig:expres02}
\end{figure}

\begin{figure}[htbp]
\centerline{\includegraphics[width=1\columnwidth]{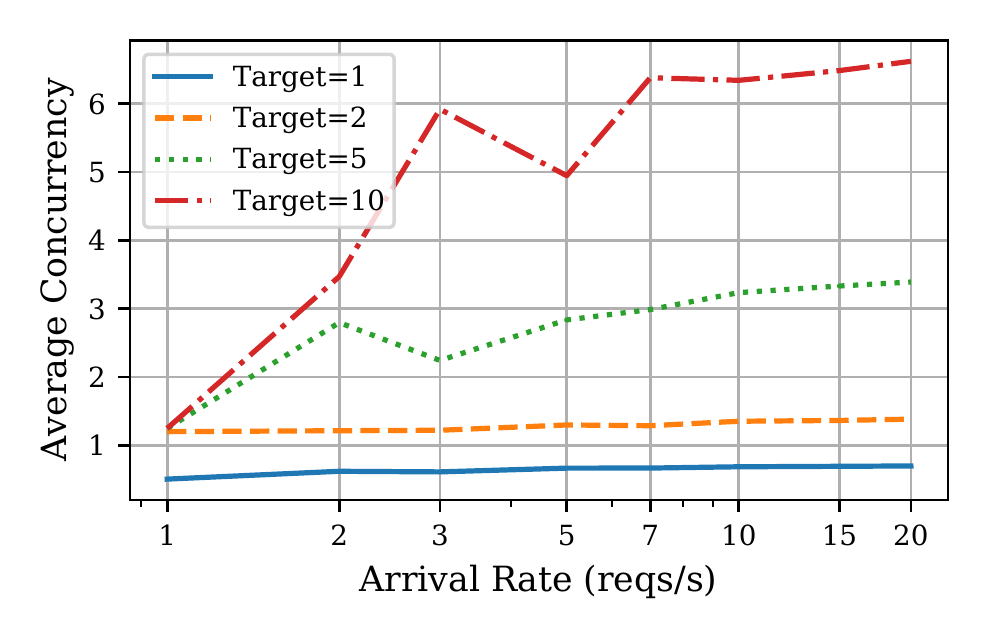}}
\caption{The predicted average concurrency versus the fixed arrival rate for different target concurrency values. Note that the x-axis is on a logarithmic scale.}
\label{fig:expres02_model}
\end{figure}

\begin{figure}[htbp]
\centerline{\includegraphics[width=1\columnwidth]{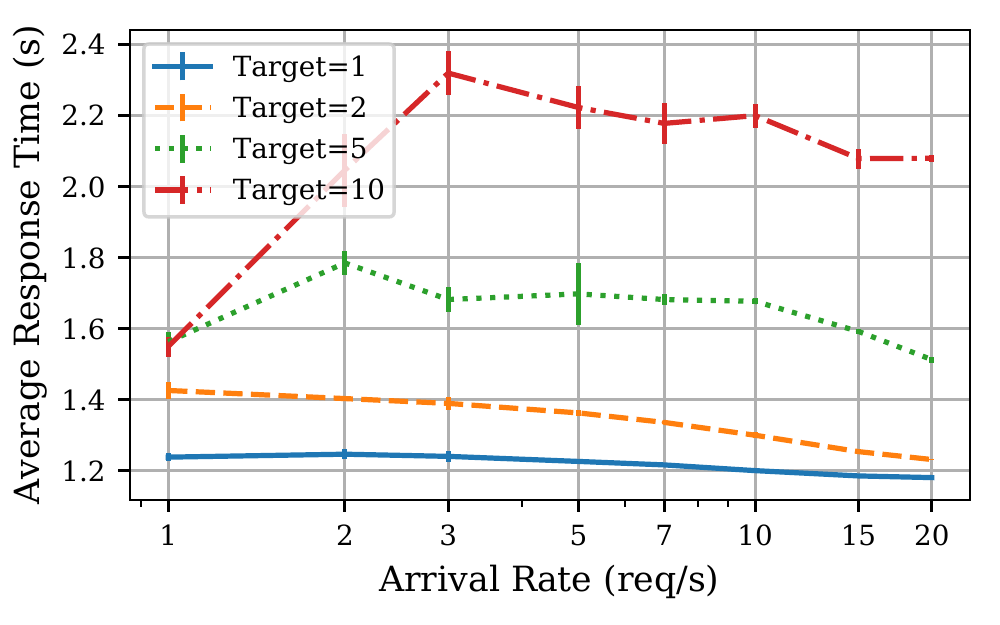}}
\caption{The measured average response time versus the fixed arrival rate for different target concurrency values in our experiments. Note that the x-axis is on a logarithmic scale. The vertical bar shows the 95\% confidence intervals.}
\label{fig:expres03}
\end{figure}

\begin{figure}[htbp]
\centerline{\includegraphics[width=1\columnwidth]{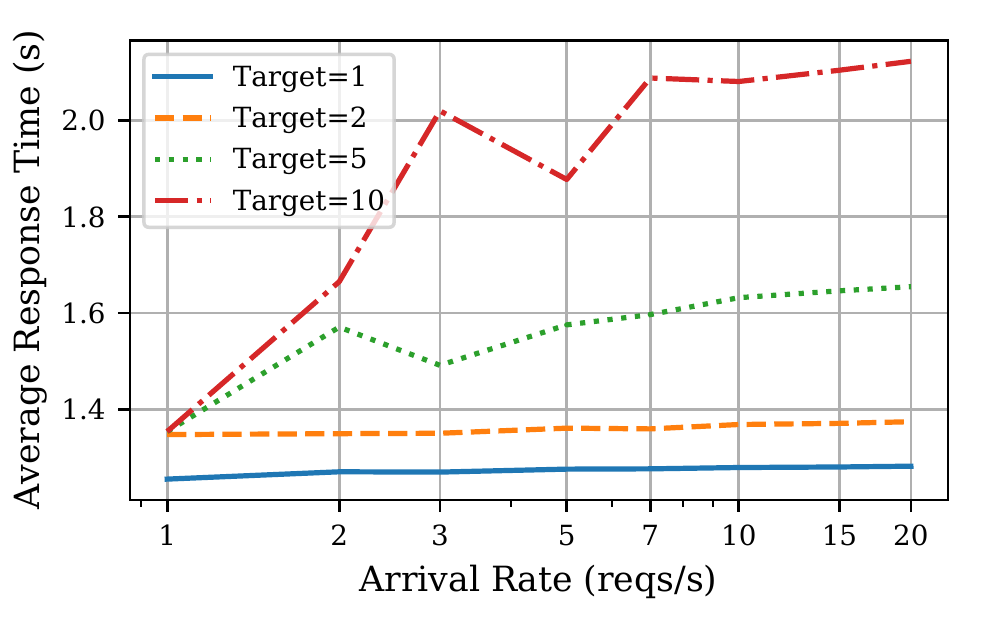}}
\caption{The predicted average response time versus the fixed arrival rate for different target concurrency values. Note that the x-axis is on a logarithmic scale.}
\label{fig:expres03_model}
\end{figure}

\Cref{fig:expres01,fig:expres01_model} show the measured and predicted average number of containers 
that are ready to serve incoming requests for different configurations, respectively. Average number
of containers are used here as a proxy to deployment cost. Depending on the setup, the deployment
cost can be VM-based in a Kubernetes cluster or Pod-based in a Google Cloud Run deployment. However,
in both scenarios, the infrastructure costs will be proportional to the average number of containers.
\Cref{fig:expres02,fig:expres02_model} depict the average concurrency value for different configurations
measured and predicted, respectively. These values can help the developer set proper configurations
for other services that the deployment relies on. For example, the provisioned capacity for most
managed database solutions can be set to optimize performance while keeping the costs low.
The average response time has been targeted here as an indicator to the deployment Quality of Service (QoS). \Cref{fig:expres03,fig:expres03_model} outline the measured and predicted average response time for
different configurations and arrival rates, respectively.
As can be seen, the experimental results shown here are well in tune with the model predictions.

\subsection{Discussion}

In \Cref{sec:exp-results}, we compared the experimental results with the performance model predictions
and showed that the effectiveness of the proposed performance model to predict the results of different
configurations for metric-based autoscaling in serverless computing platforms. The resulting performance
model can be used for any metric-based autoscaling platform as long as they adhere to the system description
outlined in \Cref{sec:system-description}. Examples of serverless computing platform that follow the
discussed system description are Google Cloud Run and Knative. To improve the tractability and accuracy
of the model while requiring a minimal amount of training data, we chose to use grey-box modelling
to integrate our knowledge about the system into the model while allowing the flexibility needed to
adapt to different types of workload.

\begin{figure}[htbp]
\centerline{\includegraphics[width=1\columnwidth]{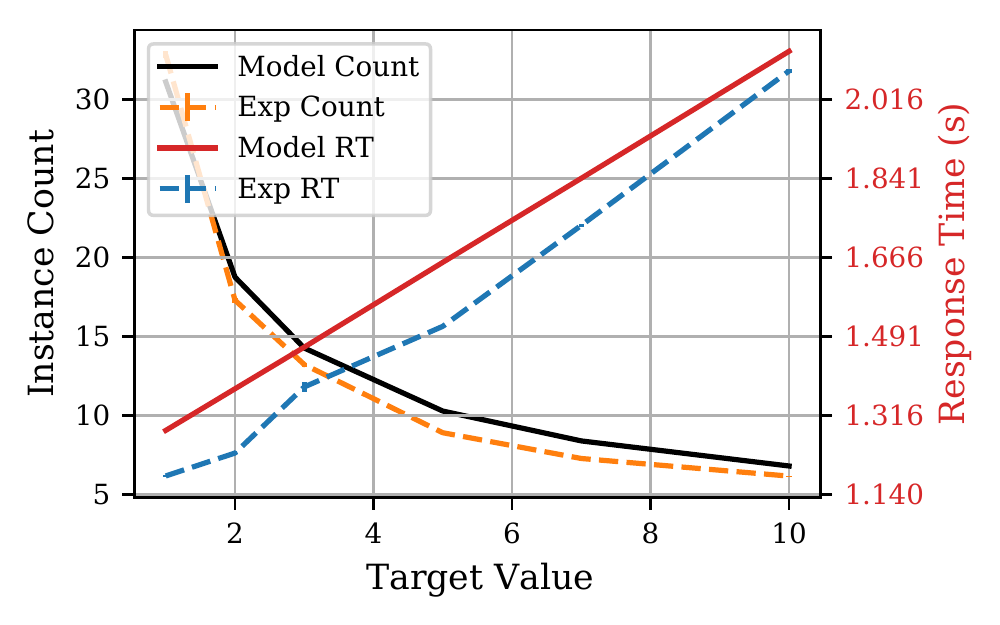}}
\caption{The effect of changing the target value on the average instance count and average response time measured in experiment and predicted by the proposed model for an arrival rate of 20 requests per second for \textit{workload 1}.}
\label{fig:expres05-arrrate_model}
\end{figure}

\begin{figure}[htbp]
\centerline{\includegraphics[width=1\columnwidth]{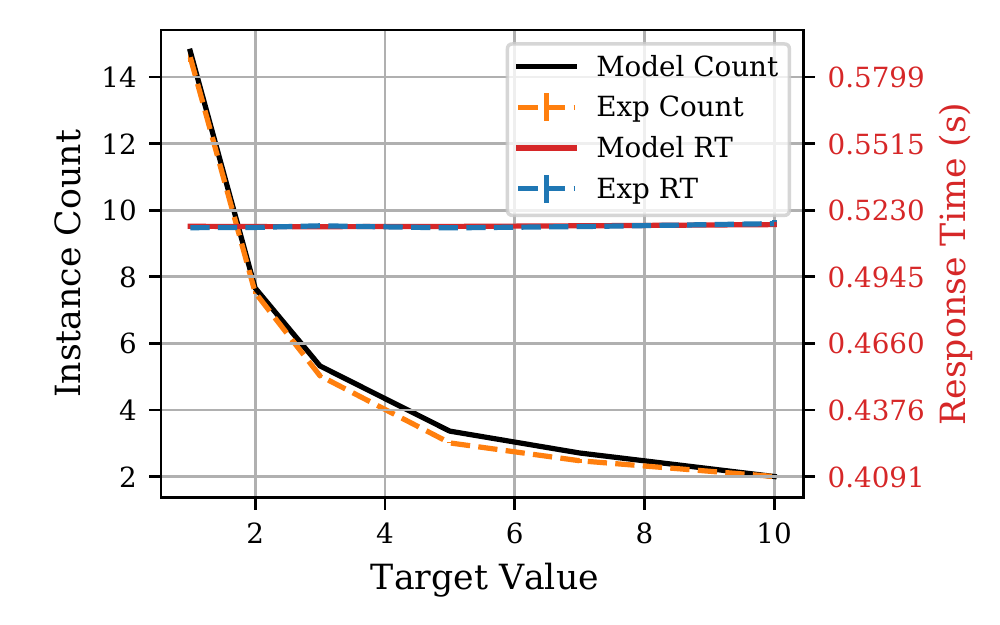}}
\caption{The effect of changing the target value on the average instance count and average response time measured in experiment and predicted by the proposed performance model for an arrival rate of 20 requests per second for \textit{workload 2}.}
\label{fig:expres05-arrrate_model-2}
\end{figure}

In \Cref{fig:expres01,fig:expres01_model,fig:expres02,fig:expres02_model,fig:expres03,fig:expres03_model},
we showed the accuracy of the proposed model in predicting key characteristics of the system under different
load intensities. By compiling these results, we can create tools that can be leveraged by the developer
to optimize their configurations by predicting the effect of a new configuration on the performance and
the cost of the system. \Cref{fig:expres05-arrrate_model} shows the measured and
predicted values for the response time and number of instances.
These figures can be used to see the effect of the target value configuration on the cost and
QoS simultaneously, which can be beneficial to make a decision about the configuration for a given
deployment. As can be seen,
the performance model can be consulted by the developer to find the optimal target value configuration
in a given system for their specific use case. As different systems have different criteria, 
finding a globally optimal point for the target value is not possible, but by presenting similar
tools, serverless providers can help facilitate a more informed decision by the developers.
\Cref{fig:expres05-arrrate_model-2} shows a similar plot but for 
\textit{workload 2}.
As can be seen, the effect of changing the chosen target value on the
quality of service varies for different workloads, but the selected regression is able
to predict this effect with sufficient accuracy.



\section{Related Work} \label{sec:related-work}

Serverless Computing has attracted a lot of attention from the research community. However, a limited number of research have focused on performance models capturing different challenges and
aspects unique to serverless computing platforms.
In previous studies, we have developed and evaluated steady-state and transient performance models
along with simulators for scale-per-request autoscaling in serverless computing platforms~\cite{mahmoudi2020tccserverless,mahmoudi2020tempperf,mahmoudi2021simfaas}.
This work is an effort to present a performance model
that captures the complexities of metric-based autoscaling, the newest trend in serverless computing
platforms, and helps us extract several important
characteristics of the serverless system.
Performance and availability have been listed on the top 10 obstacles towards the adoption of cloud services~\cite{armbrust2010view}.
Rigorous models have been leveraged to analytically model the performance of various cloud services for IaaS, PaaS, and microservices
\cite{xiong2009service,yang2009performance,khazaei2011modellingmgm,khazaei2011performance,khazaei2011performanceburst,khazaei2012fine,qian2011hierarchical,ataie2017hierarchical,chang2016effective,khazaei2020tccmicro}. 
In \cite{xiong2009service}, a cloud datacenter is modelled as a classic
open network with a single arrival. Using this modelling, the authors managed to extract the distribution
of the response time, assuming interarrival and service times are exponential. Using the response time
distribution, the maximum number of tasks and the highest level of service could
be derived. 
Yang et al.~\cite{yang2009performance} modelled the cloud datacenter as \textit{M/M/m/m+r} queuing system and derives the
distribution of response time. Assuming the periods are independent, the response time is broken down to waiting, service, and
execution later on, Khazaei et al.~\cite{khazaei2011modellingmgm,khazaei2011performance, khazaei2011performanceburst,khazaei2012fine,khazaei2020tccmicro} have
proposed monolithic and interactive submodels for IaaS cloud datacenters with enough accuracy and
tractability for large-scale cloud datacenters. Qian et al.~\cite{qian2011hierarchical} proposed a model that evaluates
the quality of experience in a cloud computing system using a hierarchical model. Their model
uses the Erlang loss model and \textit{M/M/m/K} queuing system for outbound bandwidth and response time
modelling, respectively.
Lloyd et al.~\cite{lloyd2015demystifying} developed a cost prediction model for service-oriented applications (SOAs) deployments to the cloud. Their model can be leveraged to find lower hosting costs while offering equal or better performance by using different types and counts of VMs.
In \cite{khazaei2016efficiency}, the authors proposed and validated an analytical performance model
to study the provisioning performance of microservice platforms and PaaS systems operating on top of
VM based IaaS. They used the developed model to perform what-if analysis and capacity planning
for large-scale microservices.
Eismann et al.~\cite{eismannmicroservices} demonstrated the benefits and challenges that arise in the performance testing of microservices and how to manage the unique complications that arise while doing so.



Kaviani et al.~\cite{kaviani2019towards} discusses the effectiveness of several key components
of Knative and its contribution to open-source serverless computing platforms. They found the
Knative autoscaler highly effective and mature for modern workloads.

Research has been done to investigate the performance of serverless computing platforms,
but none are offering rigorous analytical models that could be leveraged to optimize the
management of the platform.
Eyk et al.~\cite{van2018addressing} looked into the performance challenges in current
serverless computing platforms. They found the most important challenges hindering the adoption
of FaaS to be the sizable computational overhead, unreliable performance, and absence of benchmarks.
The introduction of a reliable performance model for FaaS offerings could overcome some of these
shortcomings.
Kaffes et al.~\cite{kaffes2019centralized} introduced a core-granular and centralized scheduler for serverless computing platforms. The authors argue that serverless computing platforms exhibit unique properties like burstiness, short and variable execution time, statelessness, and single-core execution. In addition, their research shows that current serverless offerings suffer from inefficient scalability, which is also confirmed by Wang et al.~\cite{wang2018peeking}.
In~\cite{bortolini2019investigating}, Bortolini et al. performed experiments on several different configurations and FaaS providers in order to find the most important factors influencing the performance and cost of current serverless platforms. They found that one of the most important factors for both performance and cost is the programming language used. In addition, they found low predictability of cost as one of the most important drawbacks of serverless computing platforms.
Lloyd et al.~\cite{lloyd2018serverless} investigated the factors influencing the performance
of serverless computing platforms.
Bardsley et al.~\cite{bardsley2018serverless} examined the performance profile of AWS Lambda
as an example of a serverless computing platform in a low-latency high-availability context.
They found that although the infrastructure is managed by the provider, and it is not visible
to the user, the solution architect and the user need a fair understanding of the underlying
concepts and infrastructure.
Pelle et al.~\cite{pelle2019towards} investigated the suitability of serverless computing platforms
(AWS Lambda, in particular) for latency-sensitive applications. Thus, the main focus in their
research was on delay characteristics of the application. Their findings showed that 
there are usually several alternatives of similar services with significantly different
performance characteristics. They found the difficulty of predicting the 
application performance for a given task, one of the major drawbacks of current serverless offerings.
Hellerstein et al.~\cite{hellerstein2018serverless}
addressed the main gaps present in the first-generation
serverless computing platforms and the anti-patterns
present in them. They showed how current implementations are
restricting distributed programming and cloud computing
innovations. The issues of no global states and the inability to address the lambda functions directly over the network are some of these issues.
Eyk et al.~\cite{van2018spec} found the most important issues surrounding the widespread 
adoption of FaaS to be sizeable overheads, unreliable performance, and new forms of 
cost-performance trade-off. In their work, they identified six performance-related challenges 
for the domain of serverless computing and proposed a roadmap for alleviating these challenges.
Zheng et al.~\cite{zheng2020autoscaling} compared the performance of OpenFaaS, Kubeless, Fission,
and Knative and found that the performance of these open-sourced serverless platforms
depends on the type of workload, the runtime implementation, and the FaaS system with the
optimal set varying case by case.

Li et al.~\cite{li2018performance} used analytical models that leverage queuing theory to optimize
the performance of composite service application jobs by tuning configurations and resource allocations. 
We believe a similar approach is possible using the presented analytical model for 
serverless computing platforms.
The new paradigm shift toward using serverless computing platforms calls for redesigning the management layer of the cloud computing platforms. To do so, Kannan et al.~\cite{kannan2019grandslam} proposed GrandSLAm, an SLA-aware runtime system that aims to improve the SLA guarantees for function-as-a-service workloads and other microservices.
Akkus et al.~\cite{akkus2018sand} used application-level sandboxing and hierarchical message buses to speed up the conventional serverless computing platforms. Their approach proved to lead to lower latency and better resource efficiency as well as more elasticity than current serverless platforms like Apache OpenWhisk.
Jia et al.~\cite{jia2021nightcore} present Nightcore, which is an efficient and scalable serverless computing framework with improved invocation latency overhead and very high invocation rate. To achieve this, they designed improved scheduling modules and introduced concurrency hints to their serverless autoscaler.
Balla et al.~\cite{balla2020adaptive} introduced Libra, an adaptive hybrid vertical/horizontal 
autoscaler on OpenFaaS trying to outperform both openfaas autoscaler and Kubernetes HPA.

\section{Threats to Validity} \label{sec:threats}

In this section, we discuss different threats to the validity of our work. We will also go over some of
the limiting assumptions that we needed to make for this study to ensure that an interested reader is 
aware of their implications in the proposed performance model.

In our experiments, we used the average response time as an indicator of the Quality of Service (QoS)
and the average instance count as an indicator of costs. These may have an impact on the results obtained
if they don't fully align with the user's use case. Analyzing every possible QoS measure and the full
billing model of all modern cloud providers is infeasible. We have selected metrics that are commonly
used in load testing experiments~\cite{eismann2020microservices}. Modern cloud-native workloads are
also billed based on their provider API usage (e.g., managed machine learning APIs) and Internet traffic.
However, we believe these costs mostly depend on the total number of requests served and thus can be
calculated without the need of a performance model.

For the presented experiments, we used two workloads in different programming languages, each comprising
several configurable benchmarks that stress different resources of the computer and represent different
types of workloads. Although experimenting with all types of workloads is not possible, the accuracy
of the performance model might differ between different programming languages. Future work should
investigate further how the knowledge can be transferred between different programming languages.
We also assumed that any external APIs used by the workload have a predictable performance that
is not affected by the amount of work applied by the studied workload. This assumption was necessary
since no performance model can consider unknown variations in an external API used by the workload.

The accuracy of the proposed model depends on the accuracy of the regression used in our metric and
output model. In our experiments, we manually ensured the quality of the resulting fit but didn't
fully investigate the extent of this relationship and how much data is required to train a regression
model with sufficient accuracy. Future studies should investigate the extent of this relationship
and how much training data is needed to ensure results have a predetermined accuracy.

In our experiments, we assumed stationarity for the workloads. This tends to hold true for
most workloads, but some workloads might violate this assumption. Especially if the
incoming request can drastically affect processing time and the incoming requests change
over time, we might see significant model drift. This effect can be mitigated by retraining
the metric model over time, but the effect has not been analyzed here and is outside the
scope of this study.

Performance experiments in the cloud always have a high degree of uncertainty due to the variable
performance perceived in cloud. Using a private academic cloud allowed us to limit the variability
of the performance, but results could vary in public clouds on shared (or burstable) CPU configurations.
To mitigate this threat, we used recommended practices to obtain and report our experimental results~\cite{eismann2020microservices}.

\section{Conclusion} \label{sec:conc}

In this work, we proposed and evaluated an accurate and tractable performance model for metric-based
autoscaling in serverless computing platforms. We analyzed the implications of different system
configurations and workload characteristics of these systems and showed the effectiveness of the
proposed model through experimental validation. We also showed how the presented performance model
can be used as a tool by application owners for finding the optimal configuration
for a given workload under different loads.
Serverless providers can also use the proposed model to adopt an adaptive and more sensible
defaults for the target value configuration. They can also leverage the performance model to
optimize the cost, performance, and energy efficiency of their system according to the
real-time arrival rate.

\section*{Acknowledgement}
This research was enabled in part by support from Sharcnet (www.sharcnet.ca) and Compute Canada (www.computecanada.ca).
We would like to thank Cybera, Alberta's not-for-profit technology accelerator, who supports this 
research through its Rapid Access Cloud services.

\bibliographystyle{ieeetr}
\bibliography{bibliography}

\begin{IEEEbiography}[{\includegraphics[width=1in,height=1.25in,clip,keepaspectratio]{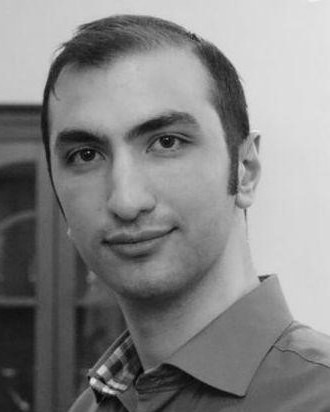}}]{Nima Mahmoudi}
received the BS degrees in Electronics and Telecommunications
and the MS degree in Digital Electronics from
Amirkabir University of Technology, Tehran, Iran in 2014, 2016, and 2017 respectively.
He is currently working towards the PhD degree in software engineering and intelligent systems at the University of Alberta, Edmonton, AB, Canada.
He is a Research Assistant at the University of Alberta and a visiting Research Assistant in the Performant and Available Computing Systems (PACS) lab at York University, Toronto, ON, Canada.
His research interests include serverless computing, cloud computing, performance modelling, applied machine learning, and distributed systems.
He is a graduate student member of the IEEE.
\end{IEEEbiography}


\begin{IEEEbiography}[{\includegraphics[width=1in,height=1.25in,clip,keepaspectratio]
{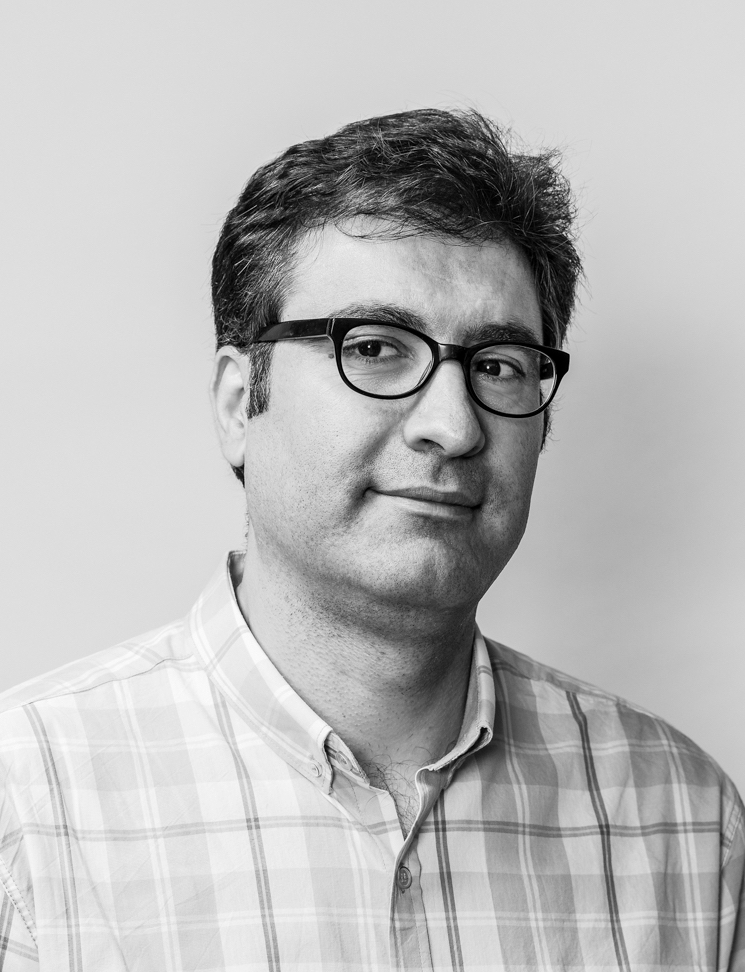}}]{Hamzeh Khazaei} (Member, IEEE) is an assistant professor in the Department of Electrical Engineering and Computer Science at York University. Previously he was an assistant professor at the University of Alberta, a research associate at the University of Toronto and a research scientist at IBM, respectively. He received his PhD degree in Computer Science from the University of Manitoba, where he extended queuing theory and stochastic processes to accurately model the performance and availability of cloud computing systems. His research interests include performance modelling, cloud computing and engineering distributed systems. 
\end{IEEEbiography}

\end{document}